\documentclass[journal=nalefd,manuscript=article]{achemso}

\usepackage[T1]{fontenc} 
\usepackage{times}
\usepackage{mathrsfs}
\usepackage{mathptmx}

\author{Boran Ma}
\affiliation[Northwestern University]
{Department of Materials Science and Engineering, Northwestern University, Evanston, IL 60208}
\author{Trung Dac Nguyen}
\affiliation[Northwestern University]
{Department of Chemical and Biological Engineering, Northwestern University, Evanston, IL 60208}
\author{Monica Olvera de la Cruz}
\email{m-olvera@northwestern.edu}
\affiliation[Northwestern University]
{Department of Materials Science and Engineering, Northwestern University, Evanston, IL 60208}
\alsoaffiliation[Northwestern University]
{Department of Chemistry, Northwestern University, Evanston, IL 60208}
\alsoaffiliation[Northwestern University]
{Department of Chemical and Biological Engineering, Northwestern University, Evanston, IL 60208}
\alsoaffiliation[Northwestern University]
{Department of Physics and Astronomy, Northwestern University, Evanston, IL 60208}

\title{Control of Ionic Mobility via Charge Size Asymmetry in Random Ionomers}


\begin{document}








\begin{abstract}
  Solid polymer electrolytes are considered a promising alternative to traditional liquid electrolytes in energy storage applications because of their good mechanical properties, and excellent thermal and chemical stability. A gap, however, still exists in understanding ion transport mechanisms and improving ion transport in solid polymer electrolytes. Therefore, it is crucial to bridge composition--structure and structure--property relationships. Here we demonstrate that size asymmetry, $\lambda$, represented by the ratio of counterion to charged monomer size, plays a key role in both the nanostructure and in the ionic dynamics. More specifically, when the nanostructure is modified by the external electric field such that the mobility cannot be described by linear response theory, two situations arise. The ionic mobility increases as $\lambda$ decreases (small counterions) in the weak electrostatics (high dielectric constant) regime. Whereas in systems with strong electrostatic interactions, ionomers with higher size symmetry ($\lambda \approx 1$) display higher ionic mobility. Moreover, ion transport is found to be dominated by the hopping of the ions and not by moving ionic clusters (also known as ``vehicular'' charge transport). These results serve as a guide for designing ion-containing polymers for ion transport related applications.
  
\end{abstract}

\section{Introduction}
\label{sec:intro}
Ion-containing polymers are a class of materials which have generated great interest over the past few decades, due to their various applications such as solid polymer electrolytes for ion batteries and fuel cells, organic field-effect transistors, and water purification membranes. \cite{hallinan2013polymer,sun2013lithium,thakur2012hybrid,kreuer2013critical,hickner2010ion} Ion transport properties in such materials are of interest in many of these applications and has been extensively investigated through experimental, theoretical, and computational approaches \cite{gomez2009effect,hall2012dynamics,Ting2015,Erbas2016,Li2016,webb2018globally,frischknecht2019evolution}. Evidently, the structural features of the assembled nanostructures, particularly the morphology of the ionic aggregates formed in ion-containing polymers, strongly influence the dynamics of co- and counterions with and without an external electric field. \cite{christie2005increasing,agrawal2015clustering} Therefore, understanding both the structure--property relationship and composition--structure relationship contributes to better design of ion-containing polymeric materials tailored for different applications. 

Charge fraction has been shown to be a very effective tuning parameter to achieve different morphologies in ion-containing polymers. \cite{sing2014electrostatic,sing2015theory,kwon2015theoretical,shim2019superlattice,park2008phase,zhou2006phase} For instance, Sing \textit{et al.} demonstrated that the Coulombic interaction between charged monomers and counterions lead to nontrivial effects in the phase behavior of polymer blends and block copolymers. \cite{sing2014electrostatic,sing2015theory} Shim and coworkers discovered a superlattice morphology by varying the charge content of charged block copolymers. \cite{shim2019superlattice} Our previous work demonstrated that ionic correlations in random ionomers lead to nanostructures with composition polydispersity, which are desired for ion battery and fuel cell applications.\cite{ma2018} Random ionomers, whose fraction and sequence of charged monomers are random, constitute an outstanding class of ion-containing polymers that is of industrial interest in that they are cost-efficiently synthesized (\textit{e.g.}, polystyrene sulfonate).

Composition heterogeneity in random copolymers of uncharged monomers results in widely different dynamic behaviors in the same system, with slow-down dynamics of frozen-like chains coexisting with Rouse-like dynamics of more mobile chains. \cite{bouchaud2019dynamics,swift1996random,semenov1999dynamics} Swift and Olvera de la Cruz demonstrated that self-diffusion of random copolymer centers of mass is substantially reduced due to the presence of quasi-frozen long chains. \cite{swift1996random} Semenov showed that there exists a cross-over molecular weight, $N^*$, beyond which the slow-down dynamics of the polymers becomes significant. \cite{semenov1999dynamics} The cross-over value $N^*$ is strongly dependent upon block polydispersity and the interaction parameter $\chi$. Microscopic-level insight into ion dynamics and transport in quenched disorder systems containing charged monomers would be beneficial for designing ionomers with desirable properties.

In previous coarse-grained simulation studies of ionomers, the size of the counterions was treated the same or very similar to that of the monomers. However, the bulkiness of the backbones of many candidate polymers for ion transport applications is key for different transport properties.\cite{griffin2018ion,paren2019impact} More importantly, given the variety of candidate counterions, how size disparity between the monomers and counterions affects the assembled morphologies and ionic dynamics remains unclear. For electrolytes, it has been shown that size asymmetry between charged particles shifts the critical temperature and critical density of vapor-liquid coexistence curve.\cite{Yan2001} Also, the suppression of the coexistence curve was attributed to larger and chain-like clusters observed in the highly size-asymmetric case.\cite{Yan2001} 

In the present study, we show that size asymmetry between monomers and counterions exhibits profound effects on the morphology of ionic clusters, ionic mobility, and transport mechanisms with and without an external electric field. Using coarse-grained molecular dynamics (MD) simulations, we predict that ion transport properties in random ionomer melts can be tuned by adjusting the size ratio between the counterions and monomers. Of particular importance is that the structural and dynamical responses of a random ionomer melt can be designed with a handful of key parameters such as degree of size asymmetry, the relative permittivity of the polymers, and the strength of the external electric field.

The degree of size asymmetry is characterized here by $\lambda$, which is the size ratio of positively charged counterions to negatively charged monomers: $\lambda = \sigma_+ / \sigma_-$, where $\sigma_+$ and $\sigma_-$ are the length scale of the Weeks-Chandler-Andersen potential\cite{weeks1971role} between the counterions and between the charged monomers, respectively. We consider four different size ratios $\lambda = 1.0, 0.5, 0.25$, and 0.1, by varying $\sigma_+$ while keeping $\sigma_{-} = 1$ as the length unit. The electrostatic interaction strength is represented by the ratio $l_B/\sigma_-$, where the Bjerrum length $l_B = q^2/(4\pi\epsilon_0\epsilon_rk_BT)$, where $\epsilon_0$ is the vacuum permittivity. The charge fraction of individual polymers, $f_C$, is sampled from a binomial distribution with a prescribed average value $\langle f_C \rangle = 0.05$, at which the charged monomers and counterions form disperse ionic aggregates for the ranges of $l_B$ of interest. \cite{ma2018} The total number density $\rho = N/V$, where $N$ is the total number of charged and uncharged monomers and counterions, and $V$ is the simulation box volume, is fixed for different size ratios. All the simulations are performed in the canonical ensemble (\textit{i.e.}, constant volume and temperature). More details are given in the Simulation Model and Method section\ref{sec:method}.

\section{Results and discussion}
\label{sec:results}
\subsection{Equilibrium properties}
\label{sec:structural}
\begin{figure}
	\centering
	\includegraphics[width=16cm]{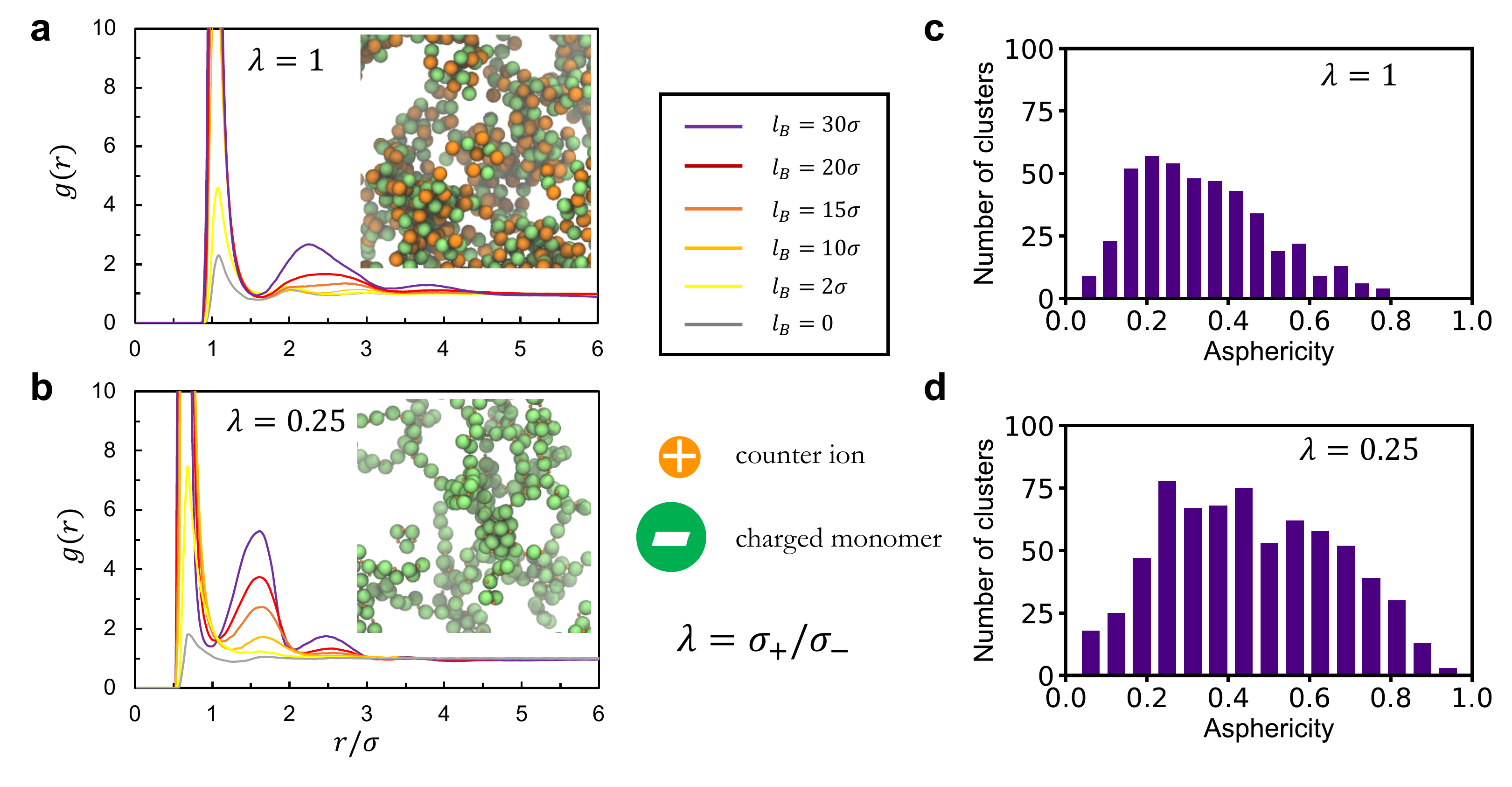}
	\caption{Pair correlation functions of negatively charged monomers and positively charged counterions in random ionomers ($f_C = 0.05$) with size ratios of (a) $\lambda = 1$, (b) $\lambda = 0.25$ (Insets: snapshots of charged species in the systems with $l_B = 30\sigma$, green beads represent charged monomers, orange beads represent counterions, neutral monomers are invisible for clarity. Snapshots are generated using VMD software.\cite{humphrey1996vmd}) and histograms of cluster asphericity in random ionomers with size ratios of (c) $\lambda = 1$, (d) $\lambda = 0.25$.}
	\label{fig:grsnap}
\end{figure}

We first characterize the microstructure formed by the charged species in random ionomers as a function of the size asymmetry and Coulombic interaction strength in the absence of an electric field. The structural properties of the ionic aggregates are investigated by the pair correlation functions between charged monomers and counterions for different values of Bjerrum lengths, $l_B$, and size disparity, $\lambda$.
It is evident from Fig. \ref{fig:grsnap}(a) and (b) that the spatial correlation between charged monomers and counterions becomes stronger with increasing $l_B$ and size disparity, as indicated by the increased height of the first and second peaks. Representative snapshots of the simulations (Fig. \ref{fig:grsnap}(a) and (b), insets) further reveal that charged species with larger size disparity (\textit{e.g.}, $\lambda = 0.25$) tend to aggregate and form more elongated clusters.

Next, we characterize the dependence of the size and shape of the clusters that are spontaneously formed by the charged monomers and counterions upon size asymmetry. We found that the average size of the ionic clusters, in terms of the number of charged particles per cluster, decreases with increasing size asymmetry for a given Bjerrum length. This can be explained by the fact that as the size of the counterions decreases relative to the charged monomers, the number of condensed counterions on the charged monomers increases, effectively screening the net attraction between the ionic clusters, thus reducing the number of charged particles per cluster. The shape of the clusters is characterized by the asphericity parameter $A_s$, which is defined as $A_s = [ (R_1-R_2)^2+(R_2-R_3)^2 + (R_3-R_2)^2]/2(R_1^2+R_2^2+R_3^2)$, where $R_1$, $R_2$, and $R_3$ are the eigenvalues of the gyration tensor of the cluster. $A_s$ ranges from  0 to 1, corresponding to a spherical shape and to an infinite rod, respectively. We show in Fig. \ref{fig:grsnap}(c) and (d) the histograms of asphericity of the ionic clusters for random ionomers with an average charge fraction of $\langle f_C \rangle = 0.05$ with different size ratios at $l_B = 30\sigma$. As $\lambda$ decreases, that is, increasing size disparity, there are more clusters formed in size-asymmetric random ionomers and the asphericity of the clusters shifts towards 1. This means that the clusters formed in size-asymmetric random ionomers are more stretched and elongated in shape. As will be shown below, these equilibrium properties still hold under a sufficiently weak external electric field.

\subsection{Responses to an external electric field}
\label{sec:efield}
The response of the ionic structures both structurally and dynamically to an external field determines the performance of polyelectrolytes in ion batteries, particularly during charging processes. Here we investigate the collective effects of the applied electric field and size asymmetry on the ionic clusters, the mobility of the counterions, and the counterion transport mechanism. The dimensionless electric field strength $E$ is defined as $E = (E_{ext}\sigma/k_BT) \sqrt{4\pi\epsilon_0\sigma k_BT}$, where $E_{ext}$ is the actual electrical field, for example, measured in the unit of V/nm.

\begin{figure}
	\centering
	\includegraphics[width=15cm]{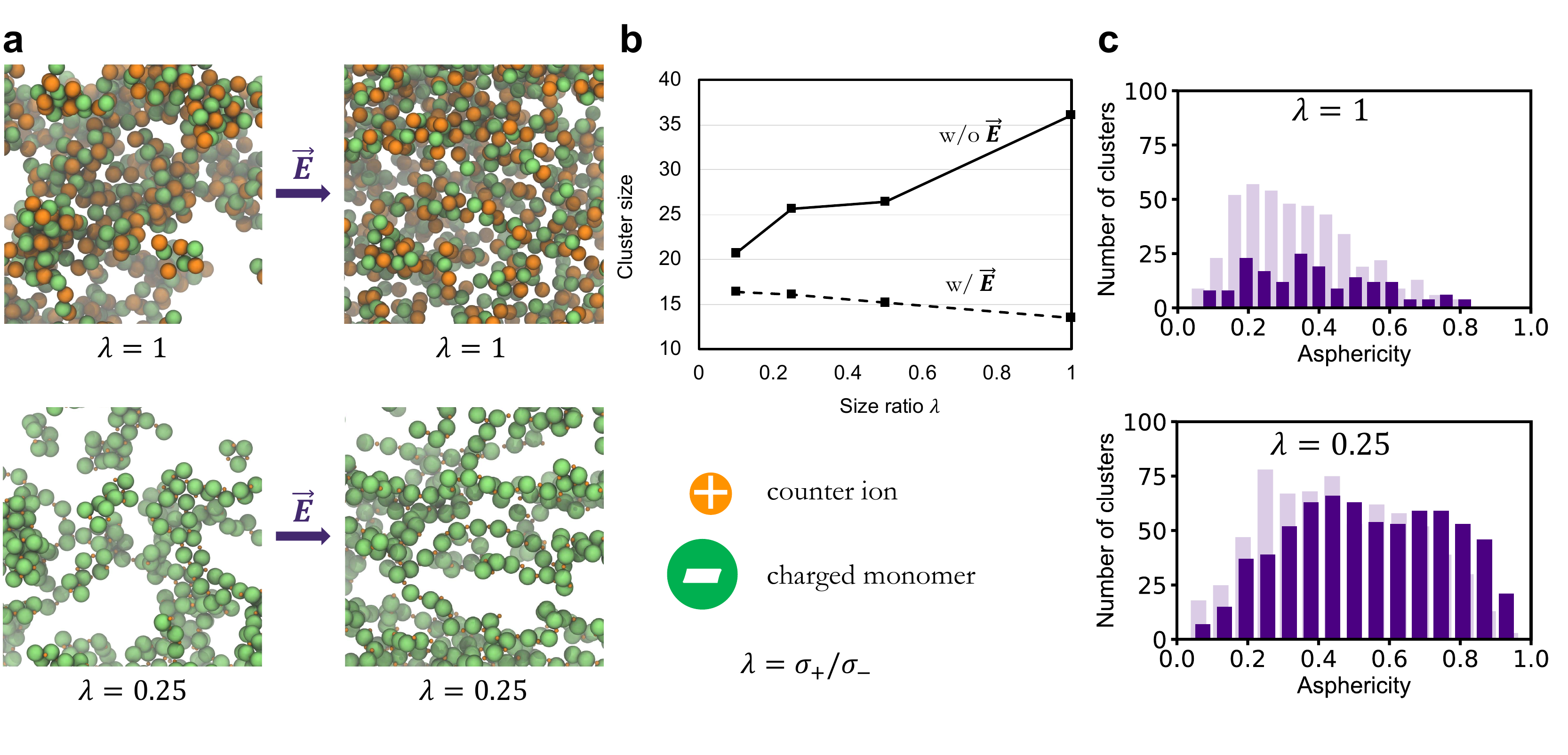}
	\caption{Morphology of charged species and geometric change of ionic clusters after a strong external electric field ($E = 10$) is applied. (a) Snapshots of charged species in random ionomers with size ratio of $\lambda = 1$ and $\lambda = 0.25$ before and after applying the external electric field. Green beads represent charged monomers, orange beads represent counterions, and neutral monomers are set to transparent for display purpose. (b) Average cluster size (in terms of number of charged particles) as a function of size ratio with and without the electric field. (c) Histograms of cluster asphericity with $\lambda = 1$ and $\lambda = 0.25$. (Light purple: without an exterinal field; dark purple: with an external field ($E = 10$).)}
	\label{fig:snapshotE}
\end{figure}

\textbf{Ionic clusters}. In the presence of a sufficiently strong electric field of $E = 10$, which corresponds to approximately 3.3 V/nm, the clusters of charged species align along the direction of the electric field in size-asymmetric random ionomers (Fig. \ref{fig:snapshotE}(a)). Comparing to the case without the external electric field, one can see that, when the external electric field is applied, the counterions and charged monomers in size-symmetric random ionomers are more scattered or distributed more evenly, with the ionic aggregates torn apart by the field (Fig. \ref{fig:snapshotE}(a)). However, the effects of the electric field on the formation of ionic clusters in size-asymmetric random ionomers are less obvious.
As shown in Fig \ref{fig:snapshotE}(b), without an external electric field, the average cluster size decreases with increasing size disparity, and size-symmetric random ionomers have the largest average cluster size. With the external electric field present, the average cluster size increases with increasing size disparity, which correlates with the highest ionic mobility in the size-symmetric random ionomers.

The influences of charged monomer-counterion size asymmetry on the assembled ionic clusters are further examined through the distribution of the cluster asphericity. For the size-symmetric case ($\lambda = 1$) the cluster asphericity distribution remains broad with smaller values upon the field application (Fig. \ref{fig:snapshotE}(c), top panel). Nonetheless, the total number of the ionic clusters is reduced substantially upon turning on the field. On the contrary, for the size-asymmetric case ($\lambda = 0.25$) the clusters are stretched when the field is applied, as indicated by $A_s$ shifting towards 1 (Fig. \ref{fig:snapshotE}(c), bottom panel). The total number of clusters meanwhile slightly decreases upon the application of the electric field. The change in the total number of clusters is in fact consistent with the change in the average cluster size described earlier. 

\textbf{Counterion mobility}. The structure--property relationship of interest is the counterion mobility as a function of an external electric field. Here we follow the procedure described in Ting \textit{et al.} \cite{Ting2015}, where the field is applied along the $x$ direction, and the Langevin thermostat is applied only to the particle velocities on the $y$ and $z$ directions. The mobility of the counterions under the external electric field, $E$, is defined as $\mu = \langle v_{ci}\rangle / (qE)$, where $\left\langle v_{ci}\right\rangle$ is the average drift velocity of counterions along the direction of the external electric field, $q$ is the counterion charge, and $\langle ...\rangle$ indicates averages over all the counterions and over the time frames after the system reached a steady state. For $E = 0$, the steady state is reached when the mean squared displacement (MSD) $\langle(\mathbf{r}_i(t)-\mathbf{r}_i(0))^2\rangle$ scales with $t$, corresponding to the diffusive regime. Counterion self-diffusion coefficient $D$, which is obtained by fitting the MSD data to a linear slope in the diffusive regime, shows a non-monotonic behavior where $D$ increases and then decreases with increasing size disparity (decreasing $\lambda$) for systems with weak electrostatic interactions ($l_B = 10\sigma$). This is presumably due to the balance between the increased attraction between charged monomers and the counterions and the decreased steric hinderance as the size of the counterions decreases. We notice that $D$ is reduced by roughly two orders of magnitude when the electrostatic interaction between the charged monomers and counterions is three times stronger, that is, with $l_B = 30\sigma$. This is interesting considering the average charge fraction of the ionomers is as low as 0.05.

\begin{figure}
	\centering
	\includegraphics[width=16cm]{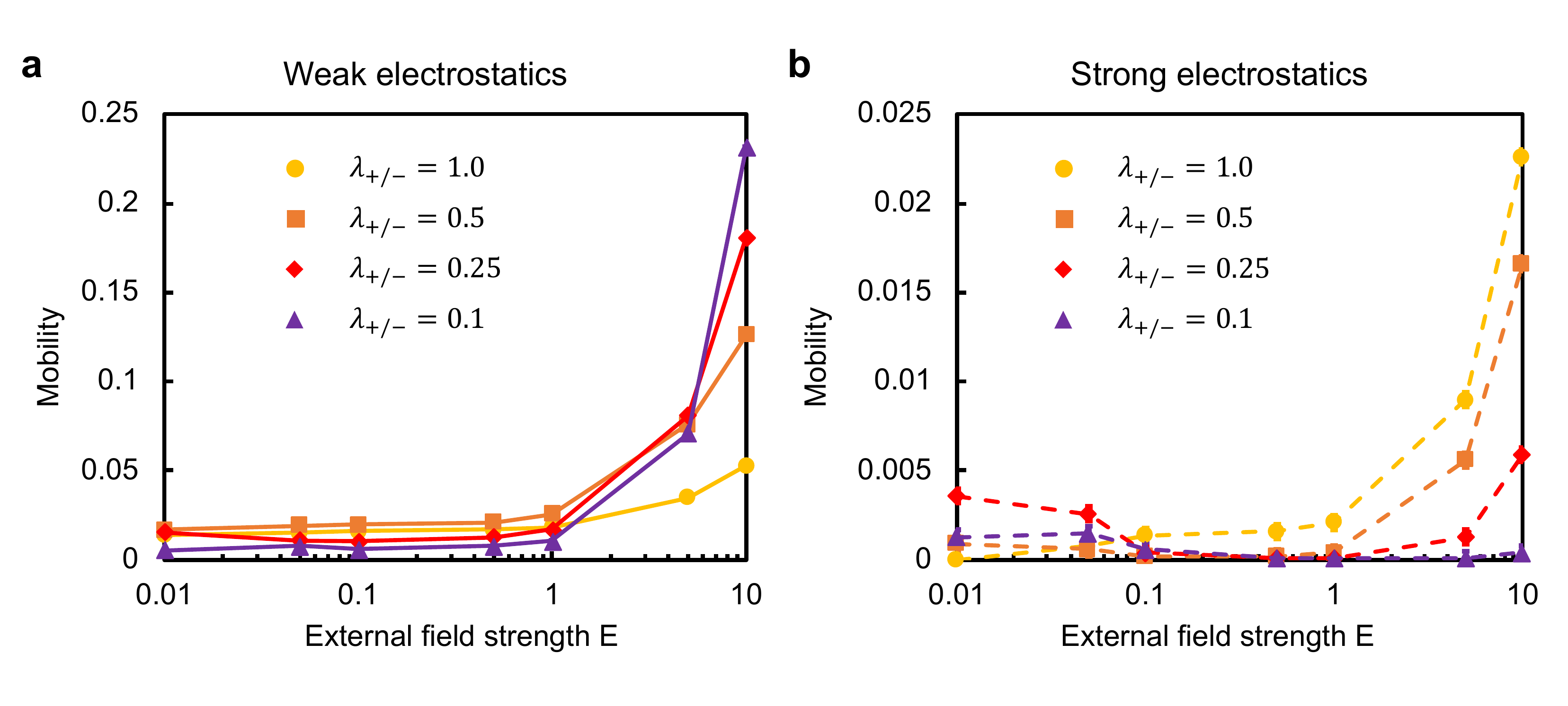}
	\caption{Mobility of counterions against external electric field strength in random ionomers with electrostatic interaction strength of (a) $l_B = 10\sigma$, and (b) $l_B = 30\sigma$.}
	\label{fig:mobility1}
\end{figure}

For $E > 0$, the steady state is reached when the mean squared displacement scales with $t^2$, that is, when the counterion velocities become constant. Ionic mobility in random ionomers with different electrostatic interaction strengths ($l_B = 10\sigma$ and $30\sigma$) under an external electric field is shown in Fig \ref{fig:mobility1}. For weak external electric field strengths ($E \leq 1$), a linear response is observed in random ionomers with both weak ($l_B = 10\sigma$) and strong ($l_B = 30\sigma$) electrostatic interactions, as indicated by the independence of the mobility upon the field strength $E$.

For stronger external electric fields (\textit{i.e.}, $E > 1$), counterion mobility increases monotonically with $E$, indicative of a non-linear response regime. As a consequence, the system is driven away from equilibrium. Note that the mobility values achieved in the weak electrostatics case are approximately an order of magnitude higher than that in the strong electrostatics case. Therefore, an ion-containing polymeric system with relatively high dielectric constant (hence a small value of $l_B$) is generally preferable for ion transport properties. This is in agreement with the findings by Wheatle et al.\cite{wheatle2018polarity} Also, in this case, increasing size disparity leads to even higher mobility, because the smaller counterions can move more freely with more free space. However, if systems with strong electrostatic interactions are chosen for trade off for other desirable properties, size symmetry is favorable, as it shows the highest mobility. This behavior can be explained by the change in the morphology of the ionic aggregates in systems with strong electrostatics with respect to the external electric field.

\textbf{Ion transport mechanism}. Efforts to reveal the ion transport mechanism in ion-containing polymers have been reported both experimentally and computationally\cite{paddison2002nature,borodin2006Li,borodin2006mechanism,castiglione2013mol,hall2012dynamics,Ting2015,Erbas2016,Li2016}The consensus from previous studies on how the counterions move in a polymer melt (in the absence of any electric field) is that the counterions have either structural diffusion or vehicular transport.\cite{paddison2002nature,borodin2006Li,borodin2006mechanism,castiglione2013mol} In the former mechanism, the counterions are hopping between oppositely charged monomers, whereas in the latter, the ions diffuse along with their neighbors, that is, an aggregate of counterions and charged monomers moving together. Various studies have concentrated on the diffusion mechanism as a function of temperature and salt concentration (see Refs. \cite{castiglione2013mol,osella2019modelling} and references therein) and in dilute polyelectrolyte solutions.\cite{fong2019ion}

In the present study, we are interested in the counterion transport mechanism in random ionomer melts under an external electric field. In the strong field limit, the system is essentially out of equilibrium. Two metrics are employed in tandem: the counterion-charged monomer decorrelation function \cite{webb2018globally} and the lifetime correlation function \cite{solano2013joint}. The counterion-charged monomer decorrelation function is defined as $h(t) = \sum_{i\in \mathcal{C}} \delta_i(t)/\sum_{i\in \mathcal{C}}\delta_i(0)$, where $\delta_i(t) = 1$ if counterion $i$ is within a cutoff distance ($r_c = 1.2\sigma$) from a certain charged monomer at time $t$, and $\delta_i(t) = 0$ otherwise. $\mathcal{C}$ is the set of the counterions that are within the cutoff distance from a charged monomer at $t = 0$. We compute the autocorrelation function of the decorrelation function $\langle h(t)h(0) \rangle$ by averaging over different time origins $t = 0$, where by definition $h(0) = 1$. Essentially, $\langle h(t)h(0) \rangle$ captures on average how long condensed counterions remain in contact with some, not necessarily the same, charged monomers. 

The lifetime correlation function is defined as $P_{ij}(t) = \sum_{i,j\in \mathcal{B}} \delta_{ij}(t) $, where $\delta_{ij}(t)=1$ if counterion $i$ is within the cutoff distance $r_c = 1.2\sigma$ from charged monomer $j$ at time $t$ and $\delta_{ij}(t) = 0$ otherwise. $\mathcal{B}$ is the set of the unique charged monomer-counterion pairs at the time origin $t = 0$. We compute the autocorrelation function of the lifetime correlation function $\langle P_{ij}(t)P_{ij}(0) \rangle$ by averaging over different time origins $t = 0$. The autocorrelation function of the lifetime correlation function shows on average how long a pair of oppositely charged particles remain in contact. The lifetime correlation function, combined with the decorrelation function, reveals how the counterions move in contact with the charged monomers. 

\begin{figure}
	\centering
	\includegraphics[width=16cm]{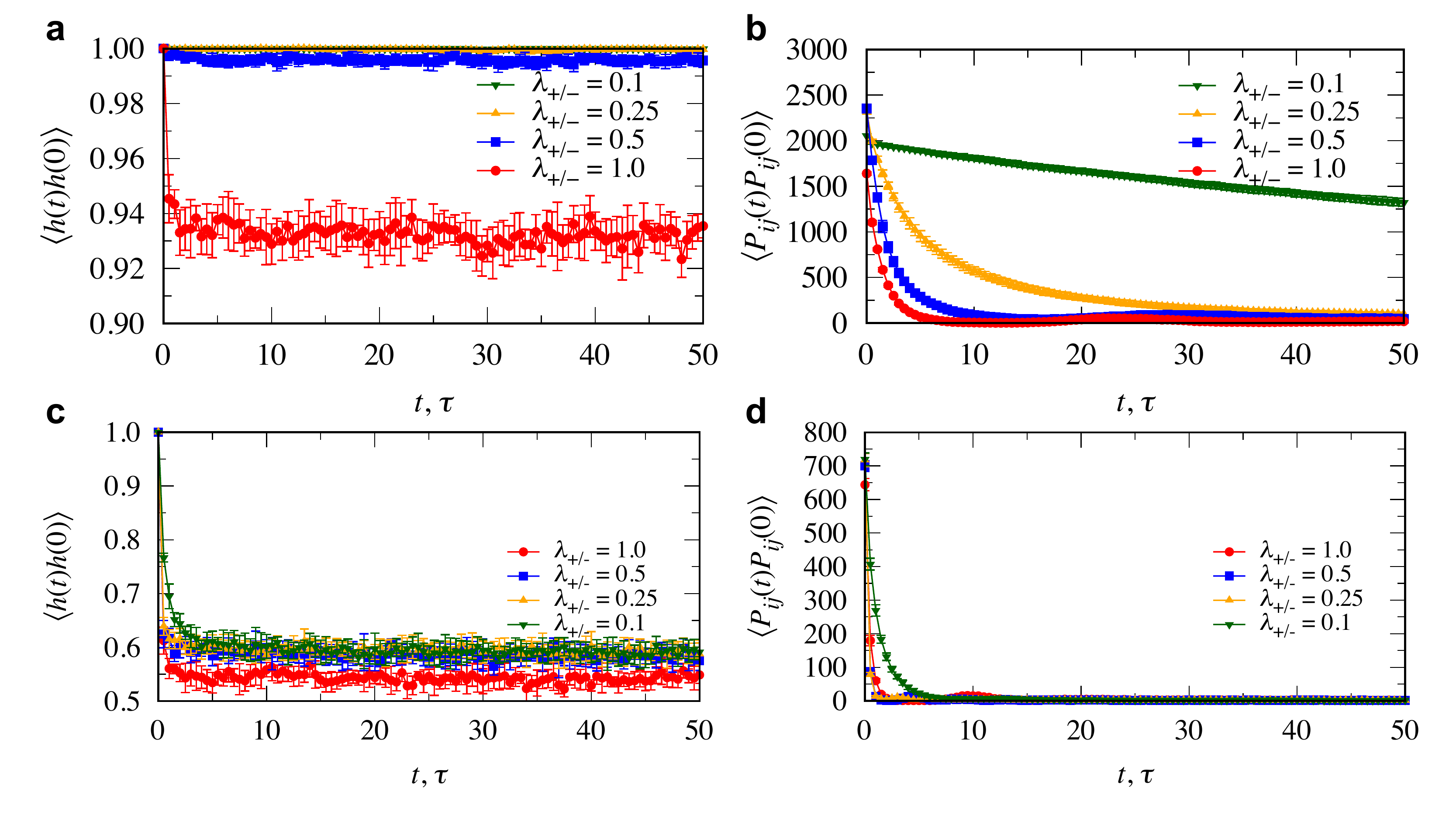}
	\caption{Interpretation of ion transport mechanisms in random ionomers under strong external electric field ($E=10$). The autocorrelation function of (a) the decorrelation function and (b) the lifetime correlation function between counterions and charged monomers in random ionomers in strong electrostatics ($l_B=30\sigma$) case; the autocorrelation function of (c) the decorrelation function and (d) the lifetime correlation function between counterions and charged monomers in random ionomers in strong electrostatics ($l_B=10\sigma$) case.}
	\label{fig:deco}
\end{figure}

For random ionomers with strong electrostatic interactions ($l_B = 30\sigma$), the autocorrelation function of the decorrelation functions (Fig \ref{fig:deco}(a)) shows that for all the studied size ratios, the majority of the counterions (greater than 90\%) are always closely bound to ionomer backbones. However, the autocorrelation function of the lifetime correlation functions for $\lambda$ = 0.25, 0.5 and 1.0 rapidly decays, indicating that in these cases the charged monomer-counterion pairs last for a very short period of time. As a result, the counterions are hopping between different charged monomers, or in other words, the ion hopping mechanism dominates in these systems. For systems with highly size-asymmetric compositions ($\lambda = 0.1$), ion hopping is suppressed, as the pairs of charged monomers and counterions remain much longer than the longest relaxation time of the polymer chains. Combined with counterion mobility for strong electrostatic interactions (Fig \ref{fig:mobility1}(b)), this indicates that hopping mechanism contributes to better ion transport properties than vehicular motion does. The ion hopping transport is also observed in systems with weak electrostatic interactions ($l_B = 10\sigma$) because almost 50\% of the counterions are not closely bound to the backbones and $\langle P_{ij}(t)P_{ij}(0) \rangle$ decay rapidly for all the size ratios studied (Fig \ref{fig:mobility1}(c) and (d)).

It should be noted that dynamics in random ionomers is expectedly highly heterogeneous, \cite{swift1996random,semenov1999dynamics} due to the fact that polymers with varying charge fraction are attached to different amounts of counterions and respond to the electric field differently. The randomness in charge fractions and sequences further gives rise to other dimensions of complexity: polymers with a higher charge fraction are expected to promote the overall dynamics and counterion mobility; whereas, those with zero charge fraction actually interfere with, or even slow down, the motion of the charged particles. Additionally, the motion of the most mobile chains and counterions should be influenced by the morphology of the assembled ionic clusters, either disperse or percolated. \cite{ma2018} 

\section{Conclusion}
\label{sec:conclusion}
We have demonstrated that size asymmetry between charged species impacts the assembled nanostructures in ionomers and can be utilized to control the ionic mobility. Different geometries of ionic clusters result from different size ratios of oppositely charged species and are further altered under the influence of an external electric field. Random ionomers with high dielectric constant are favored for ion transport, and systems with small counterions have a high ionic mobility; whereas in random ionomers with strong electrostatic interactions, counterions of comparable size as that of monomers are preferred. Correlation of motion of polymer backbones and counterions can vary by tuning size disparity, and contrary to what was found by studies in dilute solutions, the hopping mechanism dominates ionic mobility in ionomer melts. Our findings provide insights into materials design of ion-containing polymers for applications related to ion transport, for instance in fuel cell polyelectrolyte membranes, a small amount of solvent with higher dielectric constant can be added to the ionomers to achieve desired nanostructures with weak electrostatic interactions offering ion conducting pathways and high ionic mobility.






\section{Simulation Model and Method}
\label{sec:method}
A series of coarse-grained molecular dynamics (MD) simulations were performed using the HOOMD-blue software package,\cite{anderson2008general,glaser2015strong} where the random ionomers are represented by the classic bead-spring model.\cite{kremer1990dynamics} In a typical simulation, there are 500 polymer chains with degree of polymerization of $40$ plus the counterions in a cubic box with periodic boundary conditions. For polymer melts, we choose the commonly used value of the system number density $\rho = N/V = 0.85\sigma^{-3}$, where $N$ is the total number of particles and $V$ is the simulation box volume. The non-bonded excluded volume interactions between the polymer beads and between the counterions are described by Weeks-Chandler-Andersen potential\cite{weeks1971role}. For the polymer beads $\epsilon = 1$ and $\sigma_{-} = 1.0$ for both negatively charged and neutral monomers. For the positively charged counterions, $\epsilon = 1$ and $\sigma_{+} = \lambda \sigma_{-}$, where the size ratio $\lambda \le 1$. The parameters for monomer-counterion pairwise interaction are given as: $\epsilon = 1$ and $\sigma = (\sigma_{+}+\sigma_{-})/2$.

The bonds between adjacent polymer beads are modeled by finitely extensible nonlinear elastic (FENE) potential: $U_{FENE}(r) = - 0.5kR_{0}^{2}\ln[1 - (r/R_{0})^{2}]$ with $k=30.0\epsilon/\sigma^2$ and $R_{0}=1.5\sigma$ to avoid unphysical bond crossing. For each chain, the number of charged monomers is drawn from a binomial distribution with a given expected value of charge fraction $\langle f_C \rangle$. The interaction between charged particles (charged monomers and counterions) is described by the bare Coulombic interaction, and long-range electrostatics are computed with the particle-particle particle-mesh method.\cite{lebard2012self} The real space  contribution is computed with a cutoff of $r_c = 4.0\sigma$, the reciprocal space contribution is computed with a 32 by 32 by 32 mesh. The electrostatic interaction strength is represented by Bjerrum length $l_B = q^2/4\pi\epsilon_0\epsilon_rk_BT$, ranging from $0$ (the neutral case serving as a contrast) to $30\sigma$ (equivalent to a melt with a dielectric constant of $2.67$). To have different values of $l_B$ in HOOMD-Blue, we rescale the particle charge $q$ accordingly, while keeping $\epsilon_r = 1$. The systems are equilibrated in the canonical ensemble at $k_BT/\epsilon = 1$ using the Nos{\'e}-Hoover thermostat for $6\times10^7$ time steps of $0.005\tau$, where $\tau = \sigma\sqrt{m/\epsilon}$ is the reduced time unit, and then $5\times10^7$ time steps of $0.001\tau$ are carried out to investigate the impact of size asymmetry between charged species on the morphology of the ionic clusters. To study the effect of an external electric field on both the structural and dynamic properties, MD simulations were restarted from systems that previously reached equilibrium and equilibrated with a Langevin thermostat along the y and z directions (turned off in the field direction to avoid system drift due to the thermostat \cite{Ting2015}). $2\times10^7$ time steps of $0.005\tau$ are performed to gather the data for mobility and time correlation calculations described in the main text. To ascertain that the simulation results are not biased by initial configurations with strong electrostatics and/or strong field, we perform multiple runs with different random seeds used to generate initial particle velocities.

\begin{acknowledgement}

This work was supported by the U.S. Department of Commerce, National Institute of Standards and Technology as part of the Center for Hierarchical Materials Design (CHiMaD) under award no. 70NANB14H012. The authors thank the computational support of the Sherman Fairchild Foundation.

\end{acknowledgement}





\bibliography{References.bib}

\end{document}